# Spin Manipulation of Free 2-Dimensional Electrons in Si/SiGe Quantum Wells

A.M. Tyryshkin, S.A. Lyon,

*Electrical Engineering Department, Princeton University, Princeton, NJ 08544 USA*

W. Jantsch, and F. Schäffler

*Institut für Halbleiter und Festkörperphysik, Johannes-Kepler-Universität Linz, A-4040 Linz, Austria*

An important requirement for a physical embodiment of a quantum computer is that arbitrary single-qubit operations can be performed. In the case of spin-qubits, this means that arbitrary spin rotations must be possible. Here we demonstrate spin rotations of an ensemble of free 2-dimensional electrons confined to a silicon quantum well embedded in a silicon-germanium alloy host. The spins are manipulated by resonant microwave pulses and an applied magnetic field in a pulsed electron paramagnetic resonance spectrometer. From the pulsed measurements we deduce a spin coherence time in this system of about 3 µs, allowing at least 100 elementary operations before decoherence destroys the spin state. These measurements represent an important step towards the realization of quantum computation using electron spins in semiconductors, but at the same time establish some constraints on the design of such a system.

There has been considerable interest, recently, in the possibility of using the spin of an electron in a semiconductor as a qubit for quantum information processing.[1,2,3] Silicon appears to be a particularly promising host because the spin-orbit interaction is small, the dominant naturally occurring isotope has no nuclear spin, and the technology of Si integrated circuits is so well developed. It has been known for 40 years that electrons bound to donors in Si have spin population lifetimes of hours, and coherence times approaching a millisecond.[4,5,6,7] Recent results have demonstrated coherence times of at least 60 ms.[8] However, these long relaxation times are observed with isolated bound electrons in a homogeneous, bulk semiconductor, while the structures envisioned for spin-based quantum information processing require the electrons be drawn to a heterointerface and controlled by electrostatic gates.[1,2,3] A particularly promising recent proposal requires the electrons to be moved by gates to different sites at or near an interface.[9] However, there have been no reports of spin manipulation or relaxation time measurements for electrons at a heterointerface. In some compound semiconductor structures (e.g. holes at a GaAs/AlAs interface[10]) the broken symmetry at the interface can produce effective magnetic fields (Rashba fields[11]) as large as 1 T, precluding the possibility of rotating the spins by conventional means while maintaining coherence.

Here we report the first direct pulsed electron paramagnetic resonance (EPR) demonstration of arbitrary rotations of an ensemble of free-electron spins in a quantum well. While we find evidence for a Rashba field, it is sufficiently small to allow ~100 operations using conventional EPR techniques. These spin rotations represent the elementary single-qubit operations required for quantum information processing. The electrons form a high mobility 2D electron system in a one-side modulation-doped Si quantum well embedded in SiGe with the same structure as samples studied by conventional continuous wave (cw) EPR[12,13,14] and electrically detected magnetic resonance[15] (EDMR). We have utilized microwave pulses to perform the spin rotations (each rotation represents an elementary single-qubit operation required for quantum information processing) and demonstrate 2- and 3-pulse sequences to measure the spin-population lifetime or longitudinal relaxation time, $T_1$, and the coherence or transverse relaxation time, $T_2$. The decays are not purely exponential, but average relaxation times can be extracted from the data. The longest measured coherence times are $T_2 \sim 3$ μs, and the longitudinal times are $T_1 \sim 2.3$ μs. The fact that $T_2 > T_1$ is unusual and indicates that the relaxation processes are anisotropic.[16,17] The relaxation-time measurements are consistent with the suggestion, based on extensive cw EPR experiments,[14] that the longitudinal relaxation time and coherence time are controlled by an effective in-plane fluctuating magnetic field, or Rashba field, arising from the broken inversion symmetry at the heterostructure interface. We deduce a Rashba field of about 10 Gauss, orders of magnitude smaller than in some III-V systems,[10] and only observable because of the long inherent spin relaxation times in Si. In a more symmetric Si-based structure, longer spin relaxation times are to be expected.

The samples used for these experiments were grown by MBE, with a 0.5 μm strain-relaxed $Si_{0.75}Ge_{0.25}$ buffer atop a 2 μm compositionally graded $Si_{1-x}Ge_x$ layer, followed by the 20 nm Si quantum well and a 47.5 nm, modulation-doped $Si_{0.75}Ge_{0.25}$ cap. The details of the growth and electrical measurements have been described elsewhere.[18] The modulation doping density was kept low in these structures and little or no EPR signal is observed initially. However, electrons can be introduced into the quantum well by illumination and persistent photoconductivity.[12] While one can use illumination intensity and time to vary the carrier density, we have not attempted to study the pulsed EPR at different carrier densities. Instead in all of the measurements reported here we have first cooled the sample in the dark, and then illuminated it for a sufficient time to reach saturation, as measured by the cw EPR signal level. The illumination was provided by a room temperature GaAs light emitting diode with its peak emission at about 900nm, and brought to the sample through a window in the cryostat and cavity. For the structures used in these experiments, it is known that this procedure leads to a Fermi-degenerate electron system with a density of $3 \times 10^{11}/cm^2$.[13] "Annealing" the structure after illumination to 30K for a few minutes resulted in a decrease in the cw EPR linewidth (from about 150 mG to 60 mG), while leaving the integrated EPR intensity essentially unchanged.

Both the cw and pulsed measurements were performed with a Bruker Elexsys580 X-band EPR spectrometer using a dielectrically loaded cylindrical resonator (EN-4118MD4). The sample was held in a fused silica tube, immobilized with frozen ethanol, and the entire cavity and sample maintained at low temperature (4-5K) with a helium-flow cryostat (Oxford CF935). The temperature was controlled to better than 0.05K with a calibrated Cernox temperature sensor, though no temperature dependence was found over the range from 3.5 to 8K.

Two experiments were performed to demonstrate manipulation of the electron spins and to provide the information about the spin relaxation times. In both experiments microwave pulses were used to perform a single-qubit operations, rotating the spins about either the x or y-axes in a reference frame defined with the z-axis parallel to the static applied magnetic field, $B_0$, and the x- and y-axes rotating at the microwave frequency about the laboratory z-axis (A detailed explanation, including various pulse sequences, can be found in ref. 19). A 2-pulse Hahn echo experiment ($\pi/2 - \tau - \pi - \tau -$ echo) was used to measure $T_2$, which is directly available from the analysis of the echo decay as a function of interpulse delay, $\tau$. In our 2-pulse experiments the microwave frequency was offset by 7-8 MHz from the resonance line to minimize interfering background EPR signals from the resonator. This offset results in an oscillating echo shape. Both the in-phase (real) and quadrature (imaginary) components of the decaying portion of the echo signal were detected and accumulated with a transient recorder. The first ($\pi/2$) and second ($\pi$) pulse durations were 16 and 32 ns, respectively, which corresponds to an excitation bandwidth of 16 MHz (higher than the frequency offset of 7-8 MHz). Representative signals are shown in Fig. 1a for the case where the applied magnetic field is perpendicular to the 2D electron system ($\theta = 0$). A 16-step phase-cycle scheme was applied to eliminate unwanted signals, in particular the free-induction decay (FID) arising from each of the individual microwave pulses. The raw data are Fourier transformed producing an EPR line which is shifted from the carrier frequency by 7-8 MHz. In Fig. 1b we show these Fourier transformed (FT-EPR) spectra for two interpulse delays. The integrated intensity of the FT-EPR lines for different $\tau$ and $\theta = 0$ are plotted in Fig. 2a (plotted against the total time, $2\tau$) for an as-illuminated quantum well (solid squares) and after the 30K anneal (solid circles). As noted earlier, annealing produces a narrower cw EPR line, and here we find that $T_2$ is correspondingly increased. The curves shown in Fig. 2a are exponential fits to the data, giving $T_2 = 0.96$ µs and 2.98 µs before and after annealing, respectively. The fitted curves do not pass through all the data points, especially at longer delays, indicating that there is a distribution of phase memory times.

The second experiment we demonstrated was a 2-pulse inversion-recovery experiment ($\pi - T - \pi/2 -$ FID) and this was used to measure the longitudinal relaxation time, $T_1$. Similar to the $T_2$ experiment, the microwave frequency was offset by 7-8 MHz from the resonance line and the entire oscillating FID was accumulated by the transient recorder. This FID signal was Fourier transformed and the integrated intensity of the recovered EPR line was analyzed as a function of delay, T, to extract the longitudinal relaxation time $T_1$. An 8-step phase-cycle scheme was used to eliminate unwanted signals: the FID from the first microwave pulse and the echo signal originating from the two pulses. The $\pi$ and $\pi/2$ pulse durations were again 32 and 16 ns, respectively. For broad weak EPR lines the FID is short and lost in the cavity ring-down. To avoid this problem a third ($\pi$) microwave pulse was used to produce an oscillating echo, which is more easily detected because it is well separated in time from the applied microwave pulses. The resulting 3-pulse sequence is $\pi - T - \pi/2 - \tau - \pi - \tau -$ echo, with $\tau$ held constant at 100 ns. The results for $T_1$ are shown in Fig. 2b, with the solid squares from before the 30K anneal and the solid circles after the anneal. The curves in Fig. 2b are exponential fits to the data which show $T_1 = 0.6$ µs in the pre-anneal experiments and $T_1 = 2.0$ µs post-anneal ($\theta = 0$). As with $T_2$, from the quality of the fits we see that there is a distribution of spin relaxation times.

It is particularly striking that $T_2 > T_1$ in these structures. This is an unusual situation and requires the relaxation processes to be anisotropic. From an abstract perspective, any process leading to relaxation or decoherence of a spin can be viewed as a fluctuating magnetic field acting on the spin. In the Redfield limit ($\gamma \delta B \tau_c \ll 1$, where $\tau_c$ is the correlation time of the fluctuations and $\gamma$ is the electron gyromagnetic ratio) and assuming that the fluctuating fields, $\delta B$, along different spatial axes are uncorrelated, the relaxation times are given by:[17]

$$1/T_1 = g^2(\overline{dB_x^2} + \overline{dB_y^2})\frac{t_c}{1+w_0^2 t_c^2}$$

$$1/T_2 = g^2 \overline{dB_z^2} t_c + 1/2T_1$$

where the external magnetic field, $B_0$, is assumed to be applied along z direction and $\omega_0$ is the Larmor frequency of the spin in this field. If the fluctuating fields are isotropic, $T_1$ is always greater than or equal to $T_2$. On the other hand, $T_2 = 2T_1$ if $\overline{dB_z^2} = 0$ and $\overline{dB_x^2}, \overline{dB_y^2} \neq 0$.[17] This situation has been observed in other anisotropic systems.[16]

Based upon cw EPR measurements it has been suggested that the spin relaxation of these 2D electrons in Si quantum wells arises from the breaking of inversion symmetry by the Si/SiGe interface and the electric field in the quantum well.[14] This process causes the spin to feel an effective in-plane magnetic field perpendicular to the electron's wavevector, as shown by Bychkov and Rashba.[11] Such a process produces an anisotropic fluctuating field lying entirely in the plane of the 2D electron system, resulting in $\overline{dB_x^2}, \overline{dB_y^2} > \overline{dB_z^2}$ which can lead to $T_2$ being longer than $T_1$. The correlation time of the field fluctuations, $t_c$, should correspond approximately to the momentum relaxation time which is about 10ps for an electron in these high-mobility structures ($\mu \sim 90{,}000$ cm$^2$/V-sec).[13] Using this value for $t_c$ we estimate a fluctuating (root-mean-square) in-plane field of $dB_x$, $dB_y$ = 10 G and the out of plane field fluctuations to be $dB_z$ = 5 G to fit $T_1$ and $T_2$ measured after the 30K anneal. The relaxation times are shorter before the anneal, and for the same $t_c$ correspond to fields of 19 G in-plane and 8 G in the z-direction. These field values are only estimates, since it was not possible to measure the mobility of the 2D electrons at the same time and under the same conditions as the pulsed EPR, however they are consistent with the suggestion that the Rashba effect is causing the relaxation. The 30K anneal may affect the spin relaxation times in two ways: first, it could change the distribution of charges in the SiGe regions, reducing the asymmetry of the quantum well and the Rashba fields, or second, it could be changing $\tau_c$ through the mobility. It is unclear at this stage what is occurring, and further work on more symmetric structures will be necessary to make a conclusive determination.

The effective Rashba fields must always lie in the plane of the 2D electrons. We can directly test for such an anisotropy by measuring the spin relaxation times for the spins as a function of the angle of the 2D electron system with respect to the applied magnetic field. By rotating the sample so that the applied field is tilted with respect to the 2D electron layer, part of the in-plane Rashba field will now contribute to $dB_z$ (with the z-direction defined by $B_0$). Thus we expect $T_2$ to decrease while $T_1$ should increase for $\theta \neq 0$. In Fig. 3 we show $T_1$ (solid squares) and $T_2$ (solid circles) for different angles of the sample with respect to the external magnetic field. As expected $T_2$ drops considerably, while $T_1$ rises. Given the uncertainty in the correlation time, and how it might be affected by changing the orientation of the sample in the magnetic field, we have not attempted to fit the angular behavior of the relaxation times. However, the effect of rotating the sample is qualitatively consistent with what one would expect from the Rashba effect.

In summary, we demonstrated the ability to perform arbitrary single-qubit operations (spin rotations) on an ensemble of free 2D electrons in a modulation-doped Si/SiGe quantum well. We find relaxation times as long as several microseconds, allowing about 100 operations before the spins lose coherence. For the applied magnetic field perpendicular to the plane of the 2D electron system, $T_1$ is about 2.3 µs and $T_2$ is about 3 µs. However, rotating the sample to put the applied field into the plane of the 2D electrons we find that $T_1$ increases to about 3 µs while $T_2$ decreases by an order of magnitude to about 0.24 µs. The large anisotropy in $T_2$ and the fact that $T_2 > T_1$ for the field perpendicular to the 2D electron layer argue that the processes giving rise to the relaxation can be viewed as effective fluctuating magnetic fields lying in the plane of the electrons. These results are consistent with the recent suggestion that the spin relaxation is being caused by Rashba fields arising from the spin-orbit interaction and the broken inversion symmetry in the quantum well.[14] We estimate Rashba fields of about 10 – 20 G in this Si/SiGe structure, several orders of magnitude smaller than one can find in compound semiconductor heterostructures. Since the spin-orbit interaction is weak in Si, the Rashba fields are small and the spin relaxation times are long.

In addition to the demonstration of spin manipulation these measurements have other important implications for utilizing electrons at heterointerfaces for quantum computing. While we have shown that a large number of spin operations can be performed within the spin coherence time, for quantum information processing it would be preferable if even more operations were possible. There are several approaches to achieving this goal. Increasing the applied magnetic field would increase the resonant microwave frequency, allowing shorter microwave pulses (assuming that sufficient microwave power is available) and thus faster operations. The magnitude of the Rashba field is independent of the applied field, and at sufficiently high microwave frequencies ($\omega_0 t_c \gg 1$) the effect of the Rashba fields will also be suppressed and longer spin coherence times will ensue. A second approach would be to reduce the Rashba field by increasing the symmetry of the potentials seen by the electrons. The structure we have studied is one-side modulation doped, with a relatively high 2D electron density, and thus has the maximum asymmetry. From a materials standpoint, symmetric structures with comparable mobilities are more difficult to fabricate, but have been grown.[20] Such devices can be expected to give longer relaxation times.

A third approach would be to laterally confine the electrons into quantum dots and quantum wires at the interface, rather than allowing them to move freely. In the ground state of a quantum dot the electron will occupy a stationary state and the Rashba fields will change its effective g-tensor but will not affect its spin relaxation times. Excited states of the quantum dot with different orbital angular momenta will experience different effective fields and therefore different g-tensors. Transitions between states will decohere the spin. Thus, it will be important to make the quantum dots as small as possible, and work at low enough temperature to freeze the electron into the ground state. Decoherence can also be introduced into the ground state if the wavefunction is distorted by gates or other electrons, thereby mixing in excited states with higher angular momentum. In most of the spin-based quantum computing proposals, such a distortion will occur during the period when 2 spins are interacting via exchange, however it has been shown that the first-order effects can be eliminated by tailoring the time dependence of the exchange coupling.[21] It will be important to minimize other unintended time-dependent distortions of the electron wavefunctions.

The work in Princeton was supported in part by the U.S. Army Research Office and the Advanced Research and Development Activity under Contract No. DAAD19-02-1-0040 and the Defense Advanced Research Projects Agency's SPINS Program through Los Alamos National Laboratory. The work in Linz was supported by the *Fonds zur Förderung der Wissenschaftlichen Forschung*, the ÖAD and the GMe in Vienna.

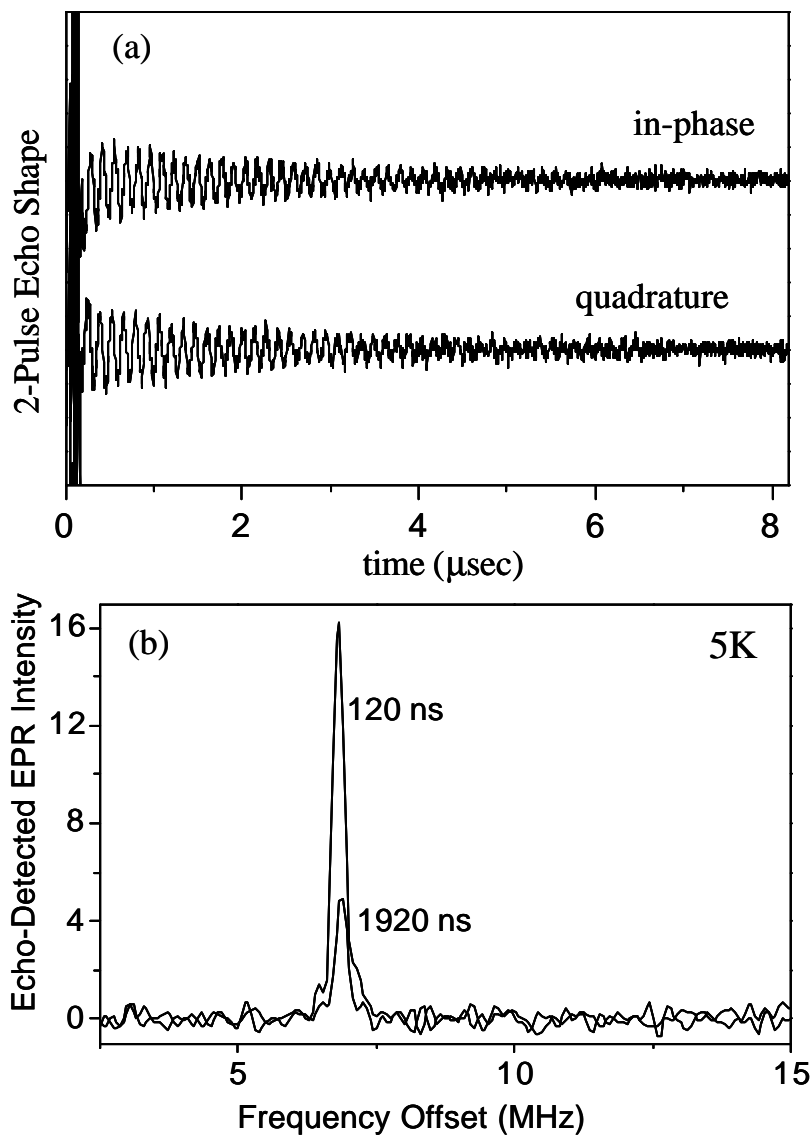

**Figure 1.** (a) Oscillating shape of the echo intensity in a 2-pulse spin-echo experiment ($\pi/2 - \tau - \pi - \tau -$ echo) after the 30K anneal of the 2D electron system with the external magnetic field, $B_0$, applied perpendicular to the plane of the electrons ($\theta = 0°$). Only the decaying portion of the echo shape is shown with the echo center at 0.1 µs (the echo rise and early decay are obscured by the cavity ring-down and detector blanking). Both the in-phase and quadrature signals are shown. (b) EPR lines obtained by Fourier transforming the decaying portion of the echo shapes for two interpulse delays, $\tau$. The frequency offset is a direct result of the oscillations in the echo shape and arises from the 7-8 MHz offset of the external magnetic field from the resonance. The integrated intensity of these FT-EPR lines is used as the measure of the echo signal intensity for determining the transverse relaxation time, $T_2$. The temperature was 5K.

Figure 2, Tyryshkin

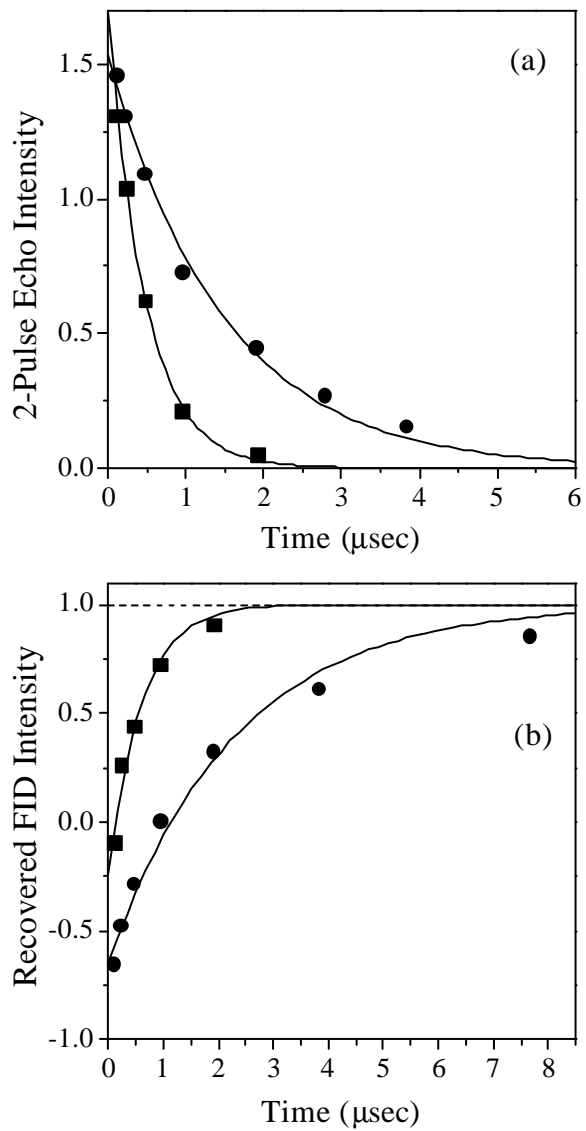

Figure 2. (a) Echo intensity in a 2-pulse experiment ($\pi/2 - \tau - \pi - \tau -$ echo) versus total delay time ($2\tau$) for as-illuminated ( ) and post 30K anneal ( ) 2D electron system. The exponential fits give $T_2$. (b) Recovered echo intensity in the inversion-recovery experiment ($\pi - T - \pi/2 -$ FID) as a function of interpulse delay T, for the as-illuminated ( ) and after 30K anneal ( ) 2D electron system. The exponential fits give $T_1$. Temperature was 4.2K and the external magnetic field, $B_0$, was applied perpendicular to the plane of the 2D electron system ($\theta = 0°$).

Figure 3, Tyryshkin

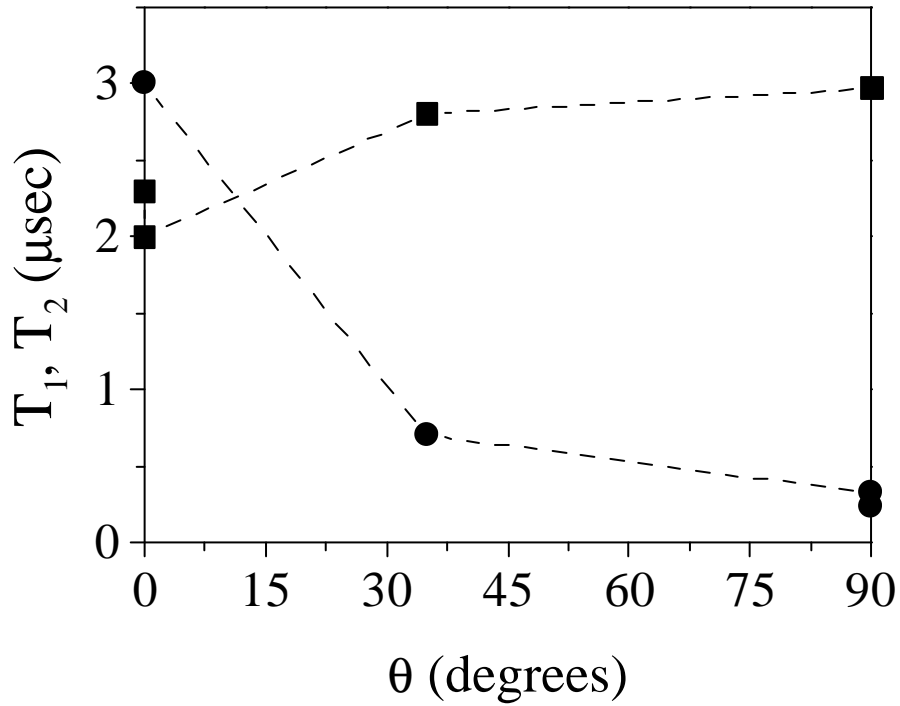

Figure 3. Electron spin relaxation times, $T_1$ ( ) and $T_2$ ( ), of the 2D electron system (after the 30K anneal) as a function of the angle ($\theta$) between the external magnetic field, $B_0$, and the normal to the 2D electron system. The lines connecting experimental points are guides to the eye. The temperature was 5K. The multiple values shown for $T_1$ at 0° and $T_2$ at 90° were obtained in different experimental runs; the differences show that the relaxation times are sensitive at the 10% level to the details of the illumination and annealing conditions. The $T_1$ points at 90° were obtained with the 3-pulse sequence ($\pi - T - \pi/2 - \tau - \pi - \tau -$ echo), as discussed in the text.